\documentclass[preprint2,numberedappendix]{emulateapj-rtx4}

\usepackage{graphicx}
\usepackage{natbib}

\usepackage{bm,times,color}
\graphicspath{{./fig/}{./png/}}
\usepackage{amsmath}

\newcommand{\EQ}{\begin{equation}}
\newcommand{\EN}{\end{equation}}
\newcommand{\EQA}{\begin{eqnarray}}
\newcommand{\ENA}{\end{eqnarray}}

\newcommand{\Eq}[1]{Equation~(\ref{#1})}

\newcommand{\Fig}[1]{Figure~\ref{#1}}

\newcommand{\Figp}[2]{Figure~\ref{#1}(#2)}

\newcommand{\Figsp}[3]{Figures~\ref{#1}({#2}) and ({#3})}

\newcommand{\Tab}[1]{Table~\ref{#1}}

\newcommand{\bra}[1]{\langle #1\rangle}

\newcommand{\meanrho}{\overline{\rho}}

{}
{}
{}

{}
{}
{}
{}
{}
{}
{}
{}
{}
{}
{}
{}
{}
{}
{}
{}
{}
{}
{}
{}
{}

{}
{}
{}

{}

{}
{}

\newcommand{\kk}{\bm{k}}

\newcommand{\BB}{\bm{B}}

\newcommand{\uu}{\mbox{\boldmath $u$} {}}

\newcommand{\JJ}{\mbox{\boldmath $J$} {}}

\newcommand{\nab}{\mbox{\boldmath $\nabla$} {}}

\newcommand{\SSSS}{\mbox{\boldmath ${\sf S}$} {}}

\newcommand{\dd}{{\rm d} {}}

\def\ga{\mathrel{\mathchoice {\vcenter{\offinterlineskip\halign{\hfil
$\displaystyle##$\hfil\cr>\cr\sim\cr}}}
{\vcenter{\offinterlineskip\halign{\hfil$\textstyle##$\hfil\cr>\cr\sim\cr}}}
{\vcenter{\offinterlineskip\halign{\hfil$\scriptstyle##$\hfil\cr>\cr\sim\cr}}}
{\vcenter{\offinterlineskip\halign{\hfil$\scriptscriptstyle##$\hfil\cr>\cr\sim\cr}}}}}
\def\Pm{\mbox{\rm Pr}_M}
\def\Rm{\mbox{\rm Re}_M}

\def\Rey{\mbox{\rm Re}}

\def\EK{E_{\rm K}}
\def\EM{E_{\rm M}}

\def\cs{c_{\rm s}}

\def\kf{k_{\rm f}}

\def\EM{E_{\rm M}}

\def\epsK{\epsilon_{\it K}}
\def\epsM{\epsilon_{\it M}}

\def\epsT{\epsilon_{\it T}}

\def\urms{u_{\rm rms}}
\def\meanurms{\overline{u}_{\rm rms}}

\def\brms{b_{\rm rms}}

\def\half{{\textstyle{1\over2}}}

\newcommand{\s}{\,{\rm s}}

\newcommand{\cm}{\,{\rm cm}}

\newcommand{\km}{\,{\rm km}}

\newcommand{\Mm}{\,{\rm Mm}}

\newcommand{\erg}{\,{\rm erg}}

\newcommand{\yjgr}[3]{ #1, {J.\ Geophys.\ Res.,} {#2}, #3}

\newcommand{\yapj}[3]{ #1, {ApJ,} {#2}, #3}

\newcommand{\yapjs}[3]{ #1, {ApJS,} {#2}, #3}

\newcommand{\yana}[3]{ #1, {A\&A,} {#2}, #3}

\newcommand{\ypf}[3]{ #1, {Phys.\ Fluids,} {#2}, #3}

\newcommand{\yprl}[3]{ #1, {Phys.\ Rev.\ Lett.,} {#2}, #3}

\newcommand{\ypnas}[3]{ #1, {Proc.\ Nat.\ Acad.\ Sci.,} {#2}, #3}

\newcommand{\yjour}[4]{ #1, {#2}, {#3}, #4}

\newcommand{\ybook}[3]{ #1, {#2} (#3)}

\begin{document}

\title{Reversed dynamo at small scales and large magnetic Prandtl number}

\author{
Axel Brandenburg$^{1,2,3,4}$\thanks{E-mail:brandenb@nordita.org}
Matthias Rempel$^{5}$
}

\affil{
$^1$Nordita, KTH Royal Institute of Technology and Stockholm University, Roslagstullsbacken 23, SE-10691 Stockholm, Sweden\\
$^2$Department of Astronomy, AlbaNova University Center, Stockholm University, SE-10691 Stockholm, Sweden\\
$^3$JILA and Laboratory for Atmospheric and Space Physics, University of Colorado, Boulder, CO 80303, USA\\
$^4$McWilliams Center for Cosmology \& Department of Physics, Carnegie Mellon University, Pittsburgh, PA 15213, USA\\
$^5$High Altitude Observatory, NCAR, P.O. Box 3000, Boulder, CO 80307, USA
}

\date{\!$ \, $Revision: 1.87 $ $\!}

\begin{abstract}
We show that at large magnetic Prandtl numbers, the Lorentz force does
work on the flow at small scales and drives fluid motions, whose energy
is dissipated viscously.
This situation is opposite to that in a normal dynamo, where
the flow does work against the Lorentz force.
We compute the spectral conversion rates between kinetic and magnetic
energies for several magnetic Prandtl numbers and show that normal (forward)
dynamo action occurs on large scales over a progressively narrower range
of wavenumbers as the magnetic Prandtl number is increased.
At higher wavenumbers, reversed dynamo action occurs, i.e., magnetic
energy is converted back into kinetic energy at small scales.
We demonstrate this in both direct numerical simulations forced by volume stirring and
in large eddy simulations of solar convectively driven small-scale dynamos.
Low density plasmas such as stellar coronae tend to have large magnetic
Prandtl numbers, i.e., the viscosity is large compared with the magnetic
diffusivity. The regime in which viscous dissipation dominates over
resistive dissipation for
large magnetic Prandtl numbers was also previously found in large eddy
simulations of the solar
corona, i.e., our findings are a more fundamental property of MHD that is
not just restricted to dynamos.
Viscous energy dissipation is a consequence of positive Lorentz
force work, which may partly correspond to particle acceleration in
close-to-collisionless plasmas.
This is, however, not modeled in the MHD approximation employed.
By contrast, resistive energy dissipation on current sheets is
expected to be unimportant in stellar coronae.
\end{abstract}


\section{Introduction}

The magnetic fields of planets, stars, accretion discs, and galaxies
are all produced and maintained by a turbulent dynamo \citep{ZRS83}.
Dynamos work through the conversion of kinetic into magnetic energy.
This energy conversion is characterized by the flow field doing work against
the Lorentz force.
It has been known for some time that this energy conversion also
depends on the microphysical value of the magnetic Prandtl number,
$\Pm\equiv\nu/\eta$, the ratio of kinematic viscosity $\nu$ to
magnetic diffusivity $\eta$ \citep{Bra09,Bra11}.
The larger the value of $\Pm$, the larger is also the ratio
of kinetic to magnetic energy dissipation \citep{Bra14}.
This is plausible, because large viscosity means large viscous
dissipation ($\epsK$), and large magnetic diffusivity or resistivity means
large resistive dissipation ($\epsM$).
Large values of $\Pm$ are generally expected to occur at low densities,
for example in the solar corona \citep[$\Pm\approx10^{10}$; see][]{Rem17}
and in galaxies \citep[$\Pm\approx10^{11}$; see][]{BS05}.

\begin{figure*}[t!]\begin{center}
\includegraphics[width=\columnwidth]{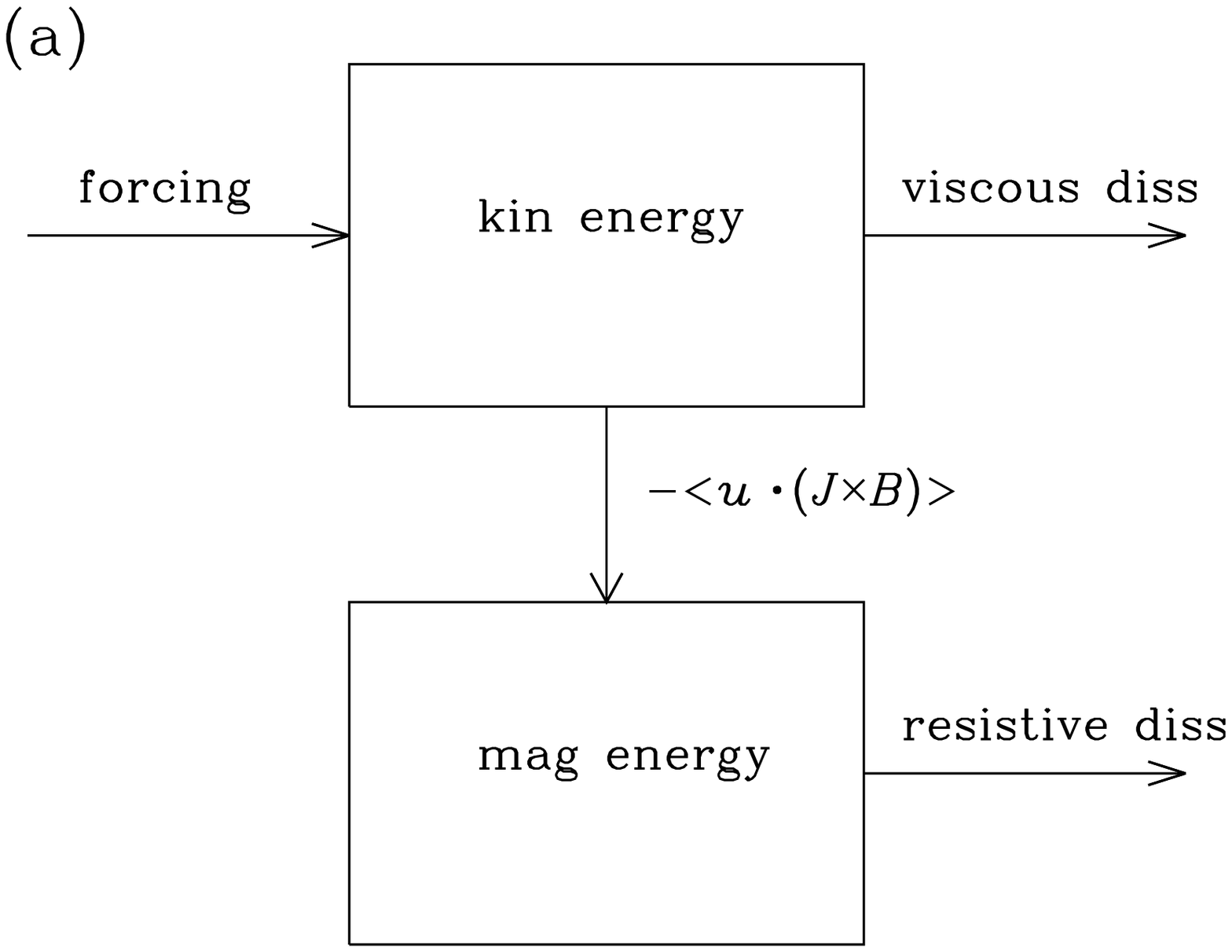}
\includegraphics[width=\columnwidth]{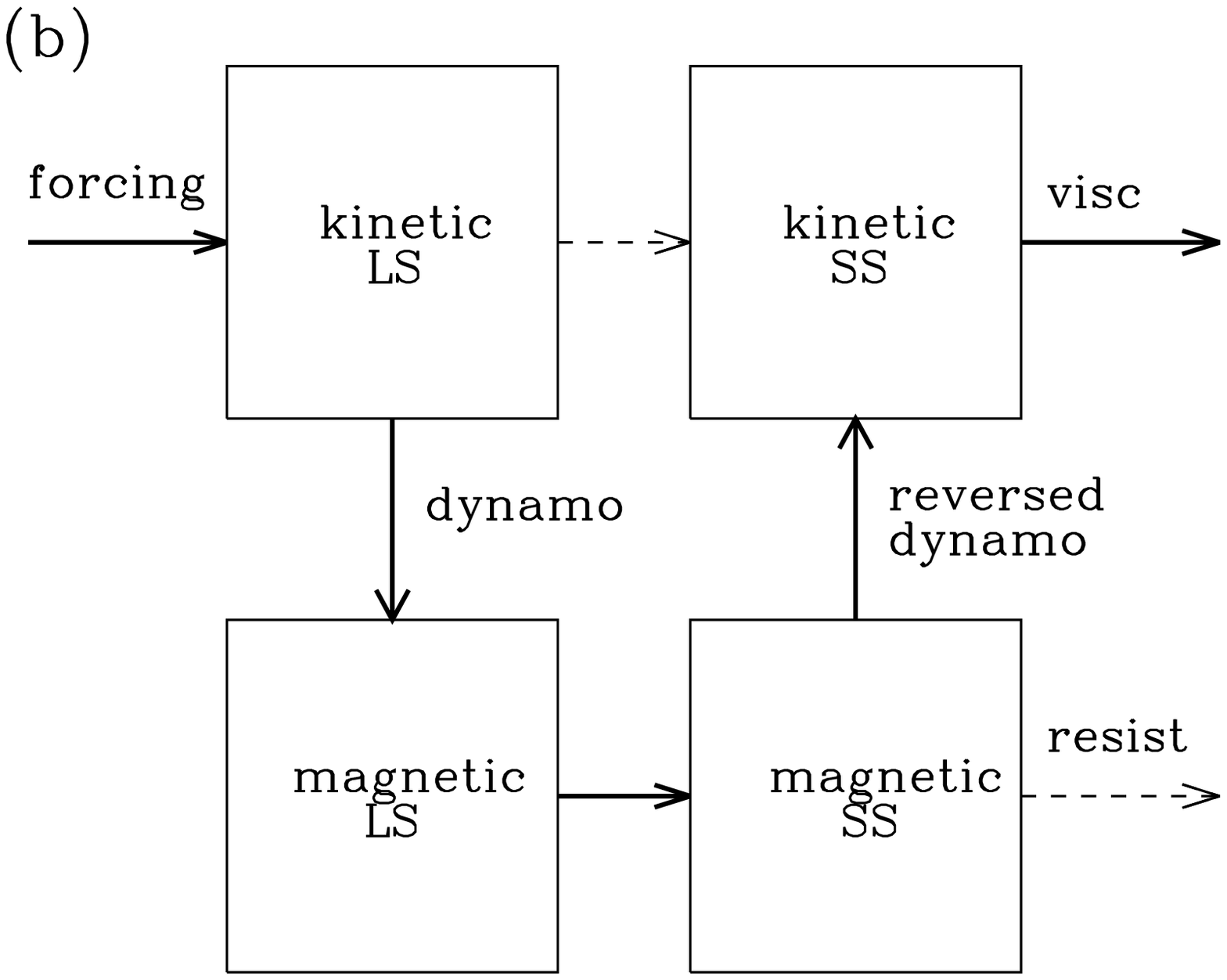}
\end{center}\caption[]{
Energy flow diagrams for (a) a standard dynamo
and (b) a dynamo at LS with a reversed dynamo at SS.
Dashed arrows indicate relatively weak flows of energy.
}\label{psketch}\end{figure*}

In the steady state, $\epsM$ must be
equal to the rate of kinetic to magnetic energy conversion.
This becomes clear when looking at an energy flow diagram; see
\Figp{psketch}{a}.
It shows that magnetic energy can only be supplied
through work done against the Lorentz force, $\JJ\times\BB$, where
$\JJ=\nab\times\BB/\mu_0$ is the current density, $\BB$ is the
magnetic field, and $\mu_0$ is the vacuum permeability.
Exactly the same amount of energy must eventually also be
dissipated resistively.
This implies that at large magnetic Prandtl numbers, not only must most
of the energy be dissipated viscously,
but also the magnetic energy dissipation must be small.
Therefore, also the work done against the Lorentz force must be
small, which suggests that the dynamo should be an inefficient one.

A large magnetic Prandtl number implies that
the magnetic diffusivity is small, so one would have expected the
dynamo to be efficient, because it suffers less dissipation.
This immediately leads to a puzzle.
How can a dynamo be efficient in the sense of experiencing low energy
dissipation, but at the same time inefficient in the sense of having
small energy conversion?

Here is where our suggestion of a reversed dynamo comes in.
A reversed dynamo is one that does work {\it by} the Lorentz force---and
not against it, as in a usual dynamo.
Thus, it corresponds to driving velocity by the Lorentz force and
hence to a conversion of magnetic to kinetic energy.
Therefore, the idea is that the flow is indeed an inefficient dynamo,
but only at large scales (LS), where kinetic energy is
converted to magnetic energy.
At small scales (SS), however, magnetic energy begins to dominate
over kinetic energy, leading therefore to an efficient conversion
of magnetic into kinetic energy.
This means we have a reversed dynamo, as sketched in \Figp{psketch}{b},
where we show the flow of energy separately for LS and SS.
To test this idea, we analyze solar convection simulations and perform
idealized simulations of isotropically forced homogeneous nonhelical
turbulence over a range of
different magnetic Prandtl numbers and calculate the spectrum of
magnetic to kinetic energy transfer.

\cite{Mahajan+05} introduced the concept of a reversed dynamo in the
context of large-scale dynamos leading to the formation of large-scale
flows that are driven simultaneously with the large-scale field by
microscopic fields and flows.
In our investigation, we focus on small-scale dynamos and show that
small-scale flows are driven by the Lorentz force when $\Pm\gg1$.
The spectral range over which this microscopic reverse dynamo is
operational is found to be $\Pm$ dependent and is largest in the high
$\Pm$ regimes.

\begin{figure*}[t]\begin{center}
\includegraphics[width=\textwidth]{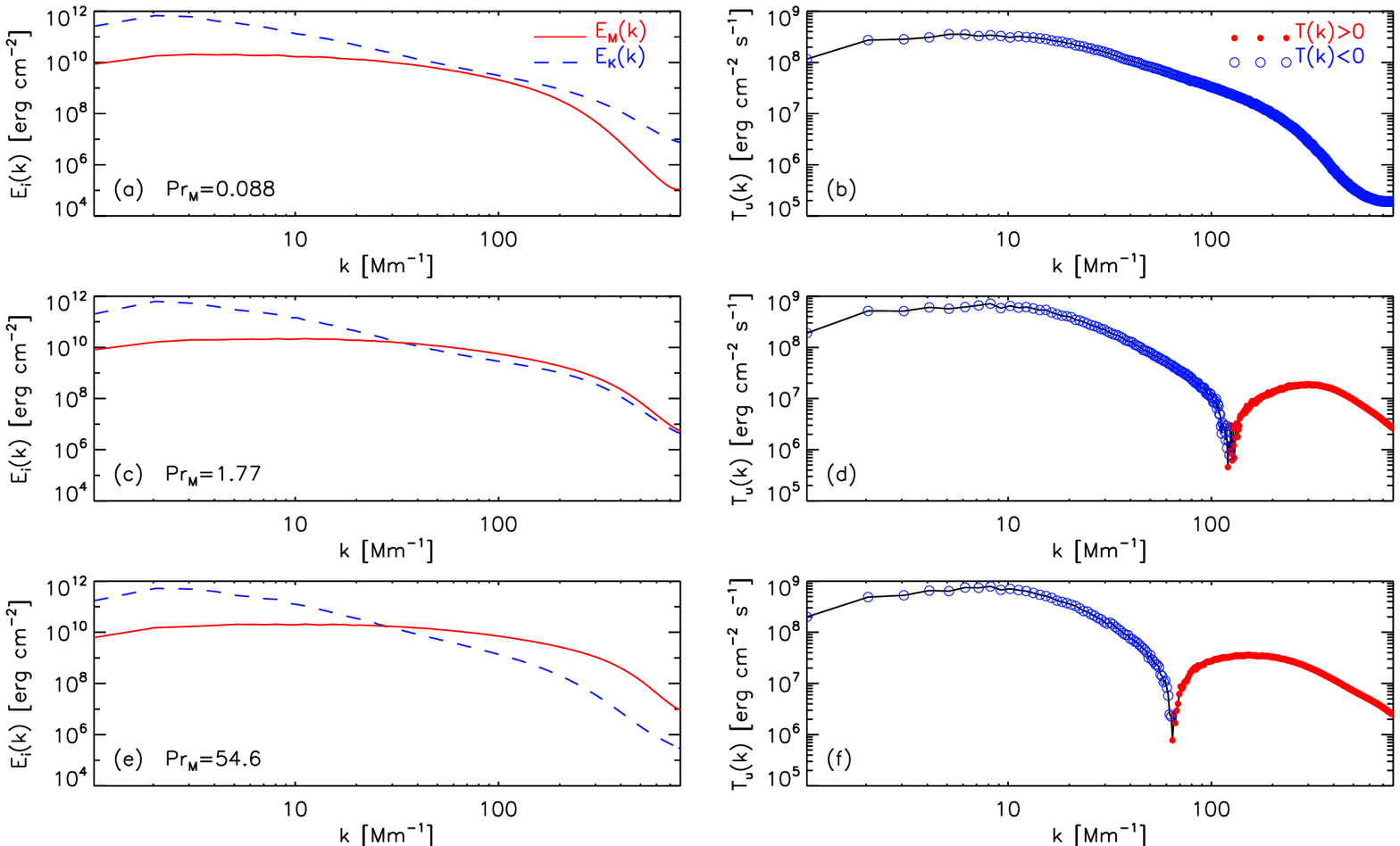}
  \end{center}\caption[]{
    $E_i(k)$ (left) and $T_u(k)$ (right) for the LES of \cite{Rem18}.
    Open blue (filled red) symbols denote negative (positive)
    values corresponding to dynamo (reversed dynamo) action.
    The values of the numerical pseudo $\Pm$ are (a,b) 0.088, (c,d) 1.77, and (e,f) 54.6.
}\label{Powerspec_Transfer_Pm}\end{figure*}

\begin{figure*}[t!]\begin{center}
\includegraphics[width=\textwidth]{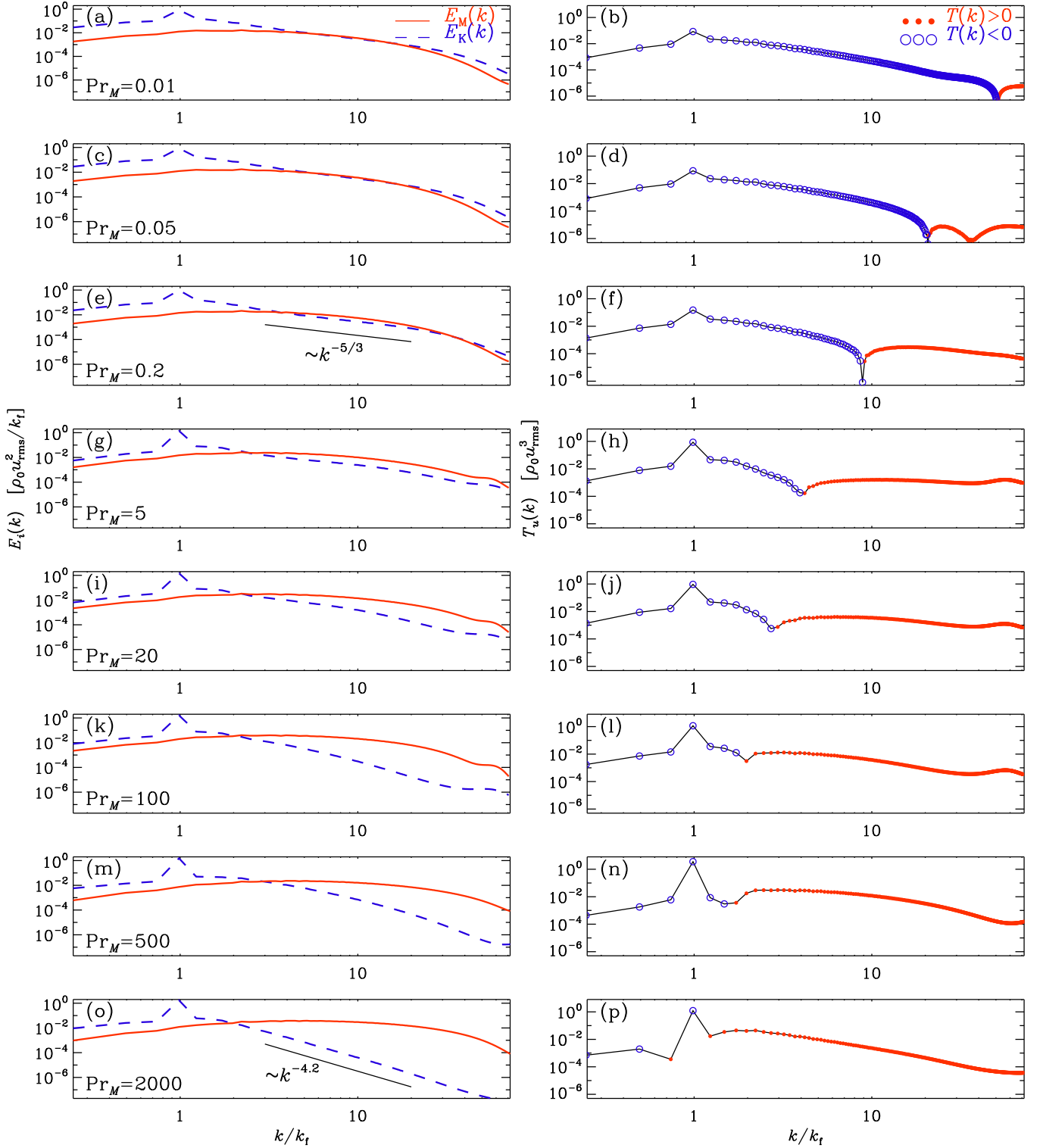}
\end{center}\caption[]{
$E_i(k)$ (left) and $T_u(k)$ (right) for the DNS of Runs~A--H (except Run~E').
Open blue (filled red) symbols denote negative (positive)
values corresponding to dynamo (reversed dynamo) action.
The values of $\Pm$ are (a,b) 0.01, (c,d) 0.05, (e,f) 0.2, (g,h) 5,
(i,j) 20, (k,l) 100, (m,n) 500, and (o,p) 2000.
The scalings of $k^{-5/3}$ (e) and $k^{-4.2}$ (o) are given for reference.
}\label{pspec_lor}\end{figure*}

\begin{figure*}[t!]\begin{center}
\includegraphics[width=\textwidth]{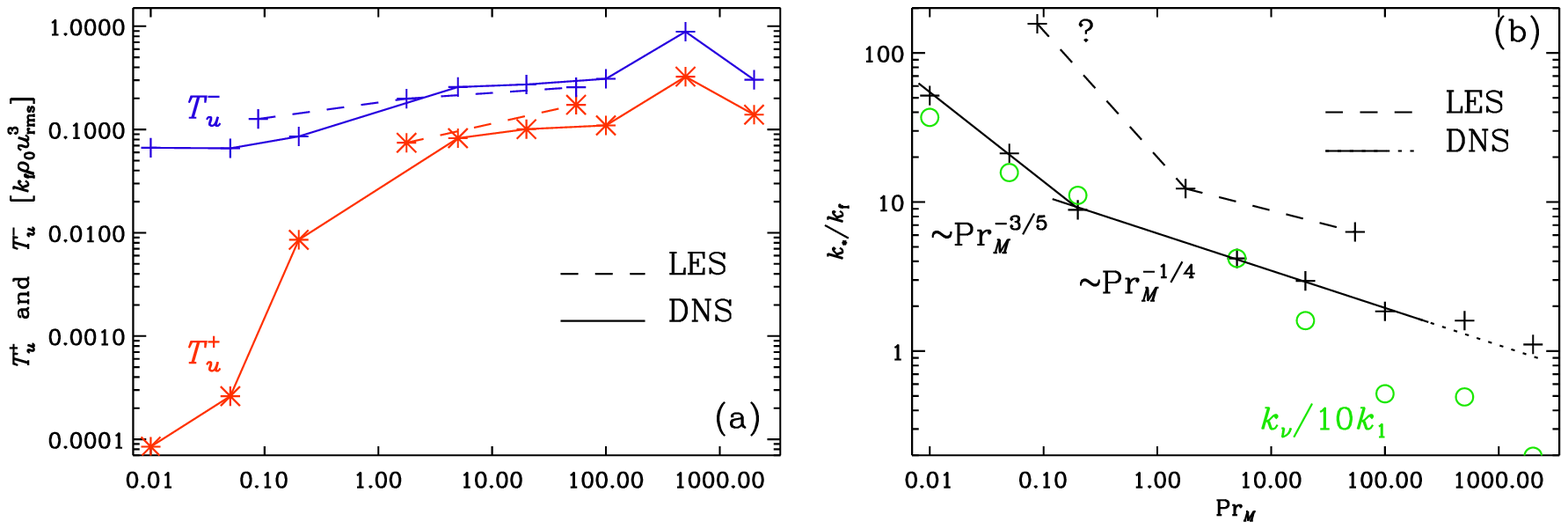}
\end{center}\caption[]{
Dependence of (a) $T_u^\pm$ and (b) $k_\ast$ on $\Pm$ for the DNS
(solid lines) and comparison with the LES (dashed lines).
In panel (b), the dotted continuation of the solid line indicates the
expected breakdown of the $\sim\Pm^{-1/4}$ scaling for $\Pm\ga200$.
The green open symbols show $k_\nu/10k_1$ for comparison.
The question mark on the leftmost LES data point indicates that $k_\ast$
is here only constrained to be $>k_{\rm max}$ for these small values
of $\Pm$.
}\label{pspec_lor_posneg}\end{figure*}

\section{Dynamo simulations and analysis}
\label{Analysis}

We consider two types of dynamo simulations.
On the one hand, we perform direct numerical simulations \citep[DNS;
as in][]{Bra14}, where viscous and magnetic dissipation are
solved for explicitly, and large eddy simulations (LES), where
these terms are modeled.
In both cases, we vary the ratio of kinetic to magnetic energy
dissipation to examine our ideas about reversed dynamo action.

Our main analysis tool is the spectrum of energy conversion, defined as
\citep{Rem14}
\begin{eqnarray}
T_u(k)=\Re\!\!\!\!\!\!\!\!\!\sum_{k_- < |{\kk}|\leq k_+} \!\!\!\!\!\!\!\!
\widetilde{\uu}_{\kk}\cdot\widetilde{(\JJ\times\BB)}_{\kk}^\ast,
\label{Evec}
\end{eqnarray}
where $k_\pm=k\pm\delta k/2$, $\delta k=2\pi/L$ is the wavenumber
increment and also the smallest wavenumber $k_1\equiv\delta k$ in the
domain of side length $L$, tildes denote Fourier transformation
either in all three directions in the homogeneous DNS or in the two
horizontal directions in the inhomogeneous LES, and $\Re$ denotes the
real part.

The LES of \cite{Rem14,Rem18} are designed to model the solar convection
zone and use realistic physics such as multifrequency radiation transport
and a realistic equation of state allowing for partial ionization effects.
Here, the flow is driven by the convection resulting from radiative
surface cooling at a rate high enough so that radiative diffusion leads
to a superadiabatic stratification, which is Schwarzschild unstable.
This model is periodic in the two horizontal directions, but not
in the vertical. The employed LES approach allows for different diffusivity
settings in the momentum and induction equations resulting in the flexibility
to change the effective numerical magnetic Prandtl number;
see Equation~(10) of \cite{Rem14}.
Beneath the solar photosphere, $\Pm$ is still well below unity,
so the purpose of changing its value is solely to demonstrate
the effect of such a change on the dynamo in the LES with the
MURaM code \citep{VSSCE05}.
However, the same code will later also be applied to a
model of the solar corona where the actual value of $\Pm$ is
much larger than unity.
Here, however, we analyze the solar convection
setups that were described in \cite{Rem18} with
effective numerical (or pseudo) $\Pm$ on the order
of $0.088$, $1.77$, and $54.6$.
It is important to note that the pseudo $\Pm=0.088$ and $1.77$ cases have
approximately the same Reynolds number, $\Rey$, but different magnetic Reynolds number, $\Rm$,
whereas the $\Pm=1.77$ and $54.6$ cases have
the same $\Rm$, but different $\Rey$.
Note that $\Pm=\Rm/\Rey$.

We should note that, owing to the strong vertical stratification
of the density $\rho$ in the LES, we have modified \Eq{Evec} by
computing $T_u(k)$ with the Fourier transforms as $\meanrho^{1/2}\uu$
and $\JJ\times\BB/\meanrho^{1/2}$, where overbars denote horizontal
averaging.
However, the choice of a $\meanrho^{1/2}$ was somewhat arbitrary
and one could have used instead $\meanrho^{1/3}$, because
$\meanrho(z)\meanurms^3(z)$ is known to be
an approximation to the convective flux which, in turn, is
expected to be approximately constant through the convection zone.
Note that the multiplication and division by the same factor
does not affect the dimension of $T_u(k)$.
Incidently, a $\meanrho^{1/3}$ factor has also been advocated by
\cite{Kritsuk} in the context of supersonic interstellar turbulence.
In the present simulations, however, $\meanrho^{1/3}\meanurms$ is seen
to increase slightly with height, while $\meanrho^{1/2}\meanurms$
decreases slightly, so the latter choice is equally well justified.

\Fig{Powerspec_Transfer_Pm} shows the
corresponding magnetic and kinetic energy spectra, $E_{\rm M}(k)$
and $E_{\rm K}(k)$, respectively, together with the spectral transfer
functions for the three cases with different pseudo $\Pm$. Since the domain is
periodic horizontally, but stratified vertically, we consider here only the horizontal
Fourier transforms when computing power spectra and transfer functions. In addition,
quantities are averaged over a height range of $800$~km ranging from $700$ to $1500$~km beneath
the solar photosphere. Unlike the DNS, the LES setup is dimensional, representing a volume of
$6.144\times 6.144\times 3.072$~Mm$^3$ in the solar photosphere.
Consequently, the LES results presented
in the following discussion are dimensional as well.

By contrast, the DNS are statistically fully isotropic and homogeneous,
so periodicity is assumed in all three directions and the flow is driven
by volume forcing with plane monochromatic nonhelical waves with random
phases and wavevectors $\kk(t)$ that are selected randomly at each time
step such that $k_-<|\kk(t)|\leq k_+$ with $k_\pm=\kf\pm\delta k/2$
and $\kf=4\,k_1$; see \cite{Bra01} for details.
We use an isothermal equation of state with constant isothermal sound
speed $\cs$. The typical Mach number based on the rms velocity varies
between $\urms/\cs\approx0.056$ for the largest value of $\Pm$ and $0.11$
for the smallest value.
We usually vary both $\nu$ and $\eta$ as we change the value of $\Pm$.

\begin{table*}
  \caption{Results from solar LES.}
  \label{tab_LES_results}
  \centerline{
    \begin{tabular}{c|c|c|c|c|c}
      Pseudo $\Pm$ & $T_u^+$ [erg cm$^{-3}$ s$^{-1}$] & $T_u^-$ [erg cm$^{-3}$s$^{-1}$] & $k_*$ [Mm$^{-1}$] & $\epsK/\epsM$ & $\meanrho\meanurms^3$ ($\bra{\rho|\uu|^3}$) $[10^{9} \erg\cm^{-2}\s^{-1}]$ \\ \hline
      $0.088$ & $0$ & $152$ & $> k_{\rm max}$ & $0.27$ & $11.5$ \quad ($19.4$) \\
      $1.77$   & $67$ & $179$ & $123$ & $0.92$ & $9.05$ \quad ($15.0$) \\
      $54.6$  & $104$ & $154$ & $63$ & $4.15$ & $5.64$ \quad ($9.06$) \\
    \end{tabular}
  }
\end{table*}

\begin{table*}\caption{
Summary of the DNS runs.
}\vspace{12pt}\centerline{\begin{tabular}{cccrrrccccccrrr}
\hline \hline
Run & $\nu k_1/\cs$ & $\eta k_1/\cs$ & $\Rey~~$ & $\Rm$ & $\Pm$ &
$\urms/\cs$ & $\brms/\cs$ & $\epsK/\epsT$ & $\epsM/\epsT$ & $\epsK/\epsM$ &
$k_\nu/k_1$ & $k_\eta/k_1$ & res.~~ \\
\hline
%
%
A  &$1\times10^{-6}$&$1\times10^{-4}$&73000&  730&   0.01& 0.110& 0.051&  0.01&  0.98&  0.01&{\em1500}& 130& $ 576^3$ \\
B  &$5\times10^{-6}$&$1\times10^{-4}$&15000&  730&   0.05& 0.110& 0.051&  0.06&  0.94&  0.07&{\em 640}& 130& $ 576^3$ \\
C  &$1\times10^{-5}$&$5\times10^{-5}$& 6800& 1400&   0.20& 0.102& 0.056&  0.13&  0.87&  0.15&{\em 450}& 220& $ 576^3$ \\
D  &$5\times10^{-5}$&$1\times10^{-5}$&  960& 4800&   5.00& 0.072& 0.051&  0.53&  0.47&  1.15& 170&{\em 540}& $ 576^3$ \\
E  &$2\times10^{-4}$&$1\times10^{-5}$&  230& 4700&  20.00& 0.070& 0.057&  0.60&  0.40&  1.51&  65&{\em 560}& $ 576^3$ \\
E' &$2\times10^{-3}$&$1\times10^{-4}$&   27&  540&  20.00& 0.081& 0.046&  0.70&  0.30&  2.31&  14& 110& $ 576^3$ \\
F  &$1\times10^{-3}$&$1\times10^{-5}$&   43& 4300& 100.00& 0.064& 0.061&  0.65&  0.35&  1.87&  21&{\em 560}& $ 576^3$ \\
G  &$1\times10^{-3}$&$2\times10^{-6}$&   31&16000& 500.00& 0.047& 0.041&  0.81&  0.19&  4.35&  20&{\em1500}& $1152^3$ \\
H  &$4\times10^{-3}$&$2\times10^{-6}$&    9&19000&2000.00& 0.056& 0.065&  0.85&  0.15&  5.85&   8&{\em1600}& $1152^3$ \\
%
%
\hline
\hline\label{Tsummary}\end{tabular}}
\tablecomments{Italics values for $k_\nu/k_1$ and $k_\eta/k_1$
are outside the reliable range.}
\end{table*}

In \Fig{pspec_lor}, we show $E_i(k)$ for $i={\rm K},{\rm M}$ and
$T_u(k)$ for eight homogeneous DNS at different
values of $\Pm$ ranging from $0.01$ to $2000$.
We normalize $T_u(k)$ by $\rho_0\urms^3$, where $\rho_0$ is the initial
(and mean) density and $\urms$ is the root-mean-square (rms) velocity.
We clearly see that at small values of $\Pm$, $T_u(k)$ is negative at almost
all values of $k$, corresponding to work done against the Lorentz work.
At larger values of $\Pm$, however, we see a progressively larger span
of wavenumbers at small scales, where $T_u(k)$ is now positive.
This corresponds to reversed dynamo action.
Similar results are also seen in the LES presented in \Fig{Powerspec_Transfer_Pm}.
For moderate values of $\Pm$, $E_{\rm M}(k)$ and $E_{\rm K}(k)$ show a
short range with $k^{-5/3}$ Kolmogorov scalings.
For large values of $\Pm$, however, $E_{\rm K}(k)$ has a much steeper
spectrum.

As $\Pm$ increases, the areas under the positive and negative parts
of $T_u(k)$,
\EQ
T_u^+=\int_{k_\ast}^\infty T_u(k)\,\dd k,\quad
T_u^-=-\int_0^{k_\ast} T_u(k)\,\dd k,
\EN
respectively, become almost equal; see \Figp{pspec_lor_posneg}{a},
where $T_u^\pm$ is normalized by $\kf\rho_0\urms^3$ for the DNS
and by $\bra{\meanrho(z)\meanurms(z)^3}$ for the LES.
However, given that there is always magnetic dissipation at the rate
$\epsM=\bra{\eta\mu_0\JJ^2}$, the 
difference between $T_u^+$ and $T_u^-$ cannot vanish
completely, but we have instead $T_u^+-T_u^-+\epsM=0$.
(We recall that $T_u^->0$ by our definition.)
The rest of the energy is dissipated viscously at the rate
$\epsK=\bra{2\rho\nu\SSSS^2}$, where $\SSSS$ is the traceless
rate-of-strain tensor; see \cite{Bra14} for details.

In both the DNS and the LES, we see that the wavenumber $k_\ast$, where
$T_u(k)$ changes sign, moves toward smaller values as $\Pm$ increases.
This is shown in \Figp{pspec_lor_posneg}{b}, where $k_\ast$ follows
an approximate $\Pm^{-1/4}$ scaling for the DNS.
For $\Pm\ga200$, however, this scaling is seen to level off.
This is expected, because in the present case the system also
has to sustain a dynamo.
For $\Pm<0.1$, however, the scaling is somewhat steeper.
Our values of $k_\ast$ are typically about ten times smaller than the
kinetic energy dissipation wavenumber, $k_\nu=(\epsK/\nu^3)^{1/4}$,
although both quantities scale similarly for $\Pm$ below five;
compare with the green open symbols in \Figp{pspec_lor_posneg}{b}.
The three data points of $k_\ast$ from the LES are also shown in
\Figp{pspec_lor_posneg}{b} and suggest a trend that is compatible
with that found in the DNS.
As seen from \Figp{Powerspec_Transfer_Pm}{b}, in the LES, $T_u(k)$
does not change sign for $\Pm<0.1$.
We can therefore only assume that $k_\ast$ is beyond
$k_{\rm max}\approx800\Mm^{-1}$.

In Table~\ref{tab_LES_results} we show the corresponding values of $T_u^+$
and $T_u^-$ for the solar LES simulations.
Similar to \Figp{pspec_lor_posneg}{b}, we find that $T_u^-$ only shows a small
dependence on $\Pm$, whereas $T_u^+$ (transfer of inverse dynamo) is increasing with $\Pm$, reducing
the net energy transfer $T_u^--T_u^+$ in the high pseudo $\Pm$ case.

It should also be noted that the sign change in the transport term
occurs only in the work done against the Lorentz force discussed above.
By contrast, magnetic energy must always be gained at all wavenumbers,
i.e., in the steady state, the corresponding transport term involving
\begin{eqnarray}
T_J(k)=\Re\!\!\!\!\!\!\!\!\!\sum_{k_- < |{\kk}|\leq k_+} \!\!\!\!\!\!\!\!
\widetilde{\JJ}_{\kk}\cdot\widetilde{(\uu\times\BB)}_{\kk}^\ast
\label{TJ}
\end{eqnarray}
must always be positive at all wavenumbers so as to balance the resistive
losses, which are proportional to $\eta|\widetilde\JJ_k|^2$ and thus
positive definite.
Looking at \Figp{psketch}{b}, this simply means that the transfer of
magnetic energy from LS to SS must be strong enough to overcome not
only the resistive losses, but also those associated with the driving
of SS kinetic energy through reversed dynamo action.
In this sense, the dynamo must be very efficient at LS.

To compare with the LES results, we normalize them
correspondingly.
In convection, $\meanrho(z)\meanurms^3(z)$ is approximately constant,
where $\meanrho(z)$ and $\meanurms(z)$, here with the $z$ argument,
indicate horizontally averaged values.
It is therefore useful to redefine
$\rho_0=\bra{\meanrho\,\meanurms^3}/\urms^3$ with
$\urms=\bra{\urms^3(z)}^{1/3}$.
From the simulations of \cite{Rem14,Rem18} we find 
$\rho_0\urms^3=(6...12)\times10^{10}\erg\cm^{-2}\s^{-1}$ for the
normalization of the spectral transfer rate adopted in \Fig{pspec_lor}.
The actual values are listed in \Tab{tab_LES_results} and compared with
the volume average of $\rho|\uu|^3$ over the height range of $800\km$
ranging from $700$ to $1500\km$ beneath the solar photosphere.
This volume average turns out to be larger than $\bra{\meanrho\,\meanurms^3}$
by a factor of between 1.6 (for large $\Pm$) and 1.7 (for small $\Pm$).
The value is comparable with the typical values seen in
\Fig{Powerspec_Transfer_Pm}.
In \Figp{pspec_lor_posneg}{b}, we compare $T_u^\pm$ from the LES,
normalized by $\kf\meanrho\,\meanurms^3\approx20\erg\cm^{-3}\s^{-1}$, where
we have used for $\kf$ the value $10\Mm^{-1}$, which is where $T_u(k)$
has a maximum; see \Figp{Powerspec_Transfer_Pm}{b}.

Interestingly, $k_\ast/\kf$ is larger in the LES than in the DNS;
see \Figp{pspec_lor_posneg}{b}.
This could well be a consequence of not having estimated $\kf$ correctly.
On the other hand, to achieve better agreement, we would need to use
a value of $\kf$ that would be even larger than $10\Mm^{-1}$, which
is already large compared with the position of the maximum of the
magnetic and kinetic energy spectra of around $2.5\Mm^{-1}$.
Another difference between LES and DNS is the complete absence of
positive values of $T_u(k)$ in the low magnetic Prandtl number LES,
which was indicated in \Tab{tab_LES_results} by writing $k_\ast>k_{\max}$.
This could be a general property of the LES which are very effective
in removing power at high wavenumbers, but it could also be an artifact
of the DNS being only marginally resolved at large values of $\Pm$.
Yet another possibility is that our vertical averaging over different
layers may have contributed to washing out the short tail at high
wavenumbers.
In either case, the trends with $\Pm$ between DNS and LES are clearly the same.

\begin{figure*}[t!]\begin{center}
\includegraphics[width=\textwidth]{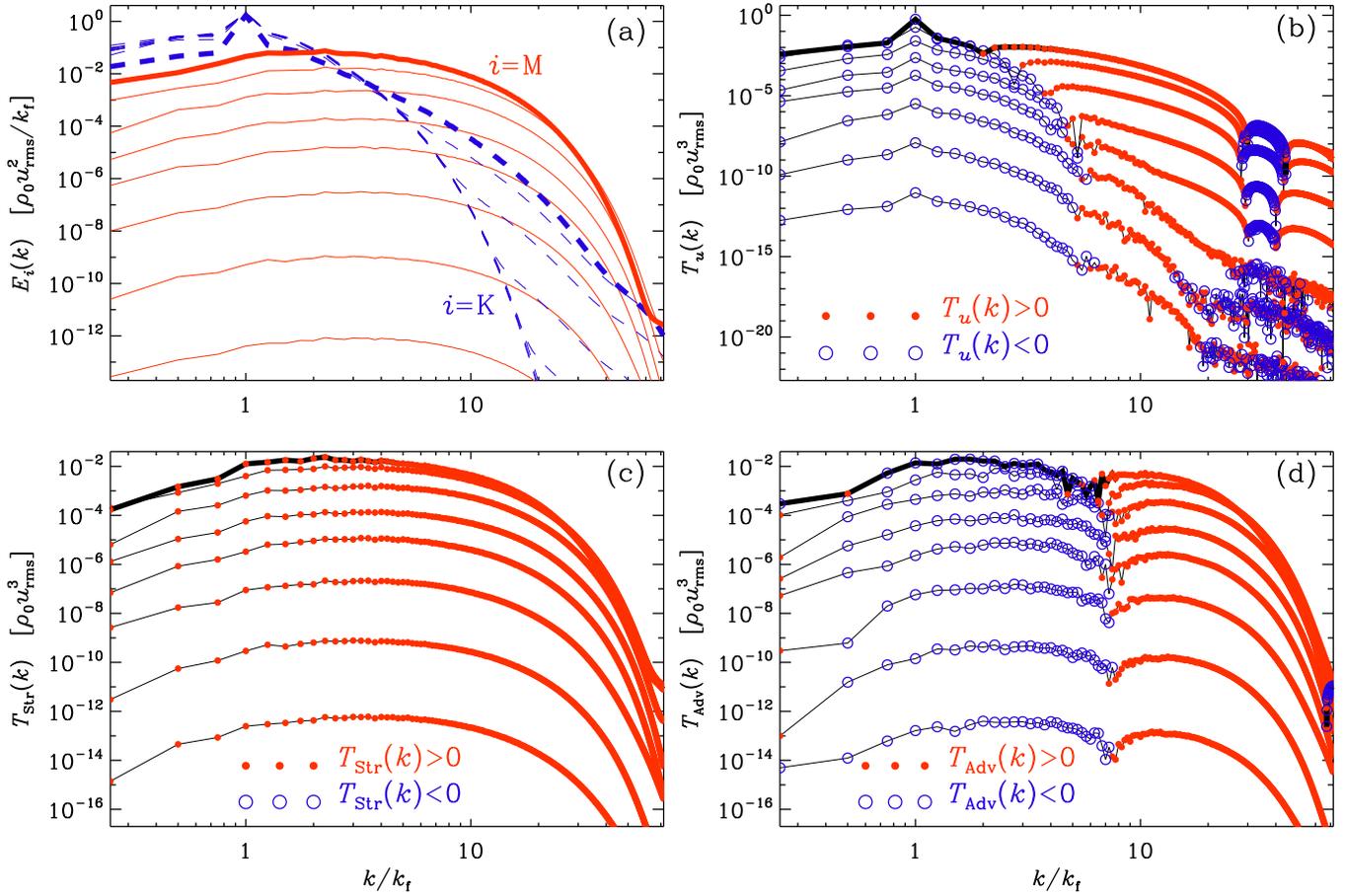}
\end{center}\caption[]{
(a) $\EK(k)$ (blue dashed) and $\EM(k)$ (red solid lines)
showing the saturation of the dynamo at eight different times
for Run~E' with $\Pm=20$.
(b) The corresponding $T_u(k)$, (c) $T_{\rm Str}$, and (d) $T_{\rm Adv}$
with filled red (open blue) symbols denoting positive (negative) values.
}\label{pspec_lor_kin}\end{figure*}

\section{Caveats}

It should be noted that the DNS at extreme magnetic Prandtl numbers
are subject to inaccuracies, whose extent cannot be fully
quantified as yet.
At unit magnetic Prandtl number, the maximum permissible Reynolds
number would never be much larger than the number of mesh points,
or, more precisely, at least for Kolmogorov-type turbulence,
the dissipation wavenumbers for kinetic
and magnetic energy dissipation, $k_\nu=(\epsK/\nu^3)^{1/4}$ and
$k_\eta=(\epsM/\eta^3)^{1/4}$, respectively, should not exceed the
Nyquist wavenumber of the mesh, $k_{\rm Ny}=\pi/\delta x$.
For $\nu/\eta\gg1$, we have $k_\eta\gg k_\nu$, and conversely for
$\nu/\eta\ll1$ we have $k_\eta\ll k_\nu$.
Thus, the larger of the two wavenumbers can severely restrict the
simulation and make it numerically unstable.
The values of $k_\nu$ and $k_\eta$ are listed in \Tab{Tsummary}
for Runs~A--H.

In the nonlinear regime, however, the rates of energy dissipation are
strongly reduced for the longer of the two spectra, because most of
the energy is dissipated through the shorter of the two spectra; see
the red solid line for $\EM(k)$ in \Figp{pspec_lor}{a} for $\Pm=0.01$
and the blue dashed line in \Figp{pspec_lor}{o} for $\Pm=2000$.
This was originally discussed in the context of small $\Pm$, where one
also has a constraint on the magnetic Reynolds number, which must be
large enough for dynamo action \citep{Bra09,Bra11}.
The simulation may then well be stable even for rather extreme
magnetic Prandtl numbers.
To what extent we can trust such simulations is unknown, but the
similarity with the LES results of \cite{Rem14,Rem18} suggests that the
opposite signs of the energy conversion spectrum and LS and
SS, as well as the change of the break point $k_\ast$ with $\Pm$ may
well be robust.
However, there can be other aspects such as the total energy conversion
rate, which may not be accurate.

\begin{figure*}[t!]\begin{center}
\includegraphics[width=\textwidth]{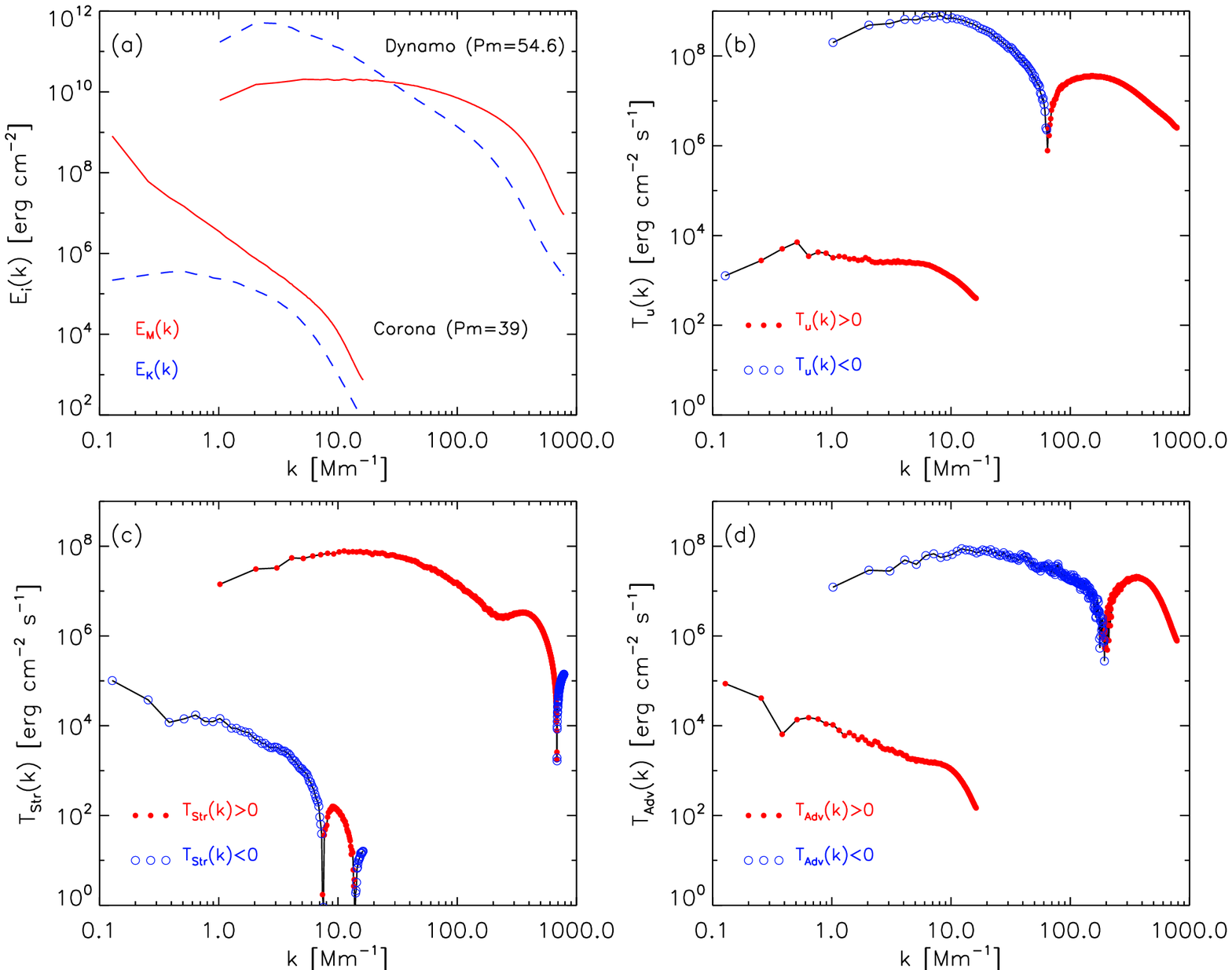}
\end{center}\caption[]{
    (a) $\EK(k)$ (blue dashed) and $\EM(k)$ (red solid lines) for a high $\Pm$
    dynamo and corona setup.
    (b) The corresponding $T_u(k)$, (c) $T_{\rm Str}$, and (d) $T_{\rm Adv}$
    with filled red (open blue) symbols denoting positive (negative) values.
}\label{Powerspec_Transfer_SSD_Corona}\end{figure*}

\section{Reversed dynamo prior to saturation}

In the kinematic phase of a dynamo, the magnetic field is too weak to
drive fluid motions.
Conversely, the field cannot resist the fluid motions.
To investigate this in more detail, we consider a simulation of a
kinematic dynamo at $\Pm=20$.
However, as explained above, in the kinematic phase it is not
safe to work with Reynolds numbers that are so large that neither
$k_\nu$ nor $k_\eta$ exceed the Nyquist wavenumber $k_{\rm Ny}$.
We have therefore computed the kinematic phase for a model
with $\Rey=27$ and $\Rm=540$, where $k_\nu=14$ or $k_\eta=110$
are well below $k_{\rm Ny}/k_1=256$.
The resulting energy spectra are shown in \Figp{pspec_lor_kin}{a}.

We see that even in this kinematic case, the same trend of
reversed dynamo action occurs in that $T_u(k)$ is positive
at small scales; see \Figp{pspec_lor_kin}{b}.
This is interesting, because \cite{Rem14} found negative values
for his kinematic pseudo $\Pm\sim 1$ simulations; see the green dashed
lines in his Figure~15(a).
This suggests that in his LES, no reversed dynamo action is
possible in the kinematic regime at $\Pm\sim 1$.
Thus, the exact location of $k_\ast/\kf$ may also be affected
by the absolute value of the fluid Reynolds number, which is
not very big in our Run~E', where also $\Pm$ is not unity.
However, as the magnetic field in Run~E' saturates, the value of
$k_\ast/\kf$ decreases from about 5 to 2.
This suggests that saturation of the dynamo is accomplished, at least
partly, by increasing the wavenumber range where reversed dynamo action
occurs.
At higher wavenumbers, $30<k/\kf<50$, there is a persistent range where
$T_u(k)$ is negative, which was not seen at larger Reynolds numbers.
It should be noted, however, that \Figp{pspec_lor}{d} gave a hint that
a sign reversal in a similar wavenumber range may be possible for
slightly different parameters.

The sign reversal of $T_u(k)$ suggests that in the subrange $30<k/\kf<50$,
magnetic energy is no longer fully sustained by the forward cascade
of magnetic energy and that it must partially by supported by forward
dynamo action just before entering the resistive subrange.
In the context of the kinetic energy spectrum in hydrodynamic turbulence,
there is a reminiscent feature known as the bottleneck, which is caused
by a sufficiently abrupt end of the inertial range \citep{Fal94}.
Whether or not such an analogy is here indeed meaningful can hopefully
be decided in future by appropriate analytic studies.

In \Figsp{pspec_lor_kin}{c}{d} we show the contributions to $T_J(k)$
from stretching and advection: $T_J(k)=T_{\rm Str}(k)+T_{\rm Adv}(k)$.
We follow here the convention of \cite{Rem14} and define
\begin{eqnarray}
T_{\rm Str}(k)&=\Re&
\!\!\!\!\!\!\!\!\!\sum_{k_- < |{\kk}|\leq k_+} \!\!\!\!\!\!\!\!\!
\widetilde{\BB}_{\kk}\!\cdot\!\left[\widetilde{+(\BB\cdot\nab\uu)}_{\kk}^\ast
-\half\widetilde{(\BB\nab\cdot\uu)}_{\kk}^\ast\right]\!\!,\;
\label{TStr} \\
T_{\rm Adv}(k)&=\Re&
\!\!\!\!\!\!\!\!\!\sum_{k_- < |{\kk}|\leq k_+} \!\!\!\!\!\!\!\!\!
\widetilde{\BB}_{\kk}\!\cdot\!\left[\widetilde{-(\uu\cdot\nab\BB)}_{\kk}^\ast
-\half\widetilde{(\BB\nab\cdot\uu)}_{\kk}^\ast\right]\!\!.\;
\label{TAdv}
\end{eqnarray}
An important reason for including one half of the compression term,
$\BB\nab\cdot\uu$, in both the stretching and advection terms is that the
energy contribution from the latter is a divergence term that vanishes
for periodic boundary conditions under volume averaging, that is,
$\int T_{\rm Adv}(k)\,\dd k=0$.
The energy contribution from the former also contains a divergence term,
which vanishes for periodic boundaries, while the remaining contribution
equals the work against the Lorentz force; see Equations~(23) and (24)
of \cite{Rem14}.
Therefore, for our DNS with triply periodic boundaries, we have
$\int T_{\rm Str}(k)\,\dd k=\int T_u(k)\,\dd k$.

Looking at \Figsp{pspec_lor_kin}{c}{d},
we see that $T_{\rm Str}(k)$ is positive for all $k$ and
$T_{\rm Adv}(k)$ is negative at small $k$ and positive at large $k$.
This agrees with the results of \cite{Rem14}, who interpreted the
negative (positive) sign of $T_{\rm Adv}(k)$ at small (large) $k$
as evidence for a transport of magnetic energy from LS
(where it acts as a loss) to SS (where it acts as a source).
This corresponds to the lower arrow between ``magnetic LS'' and
``magnetic SS'' in  \Figp{psketch}{b}.
Note that the break point, where $T_{\rm Adv}(k)$ changes sign, does not
change much as the dynamo saturates.

\section{Applications to stellar coronae}

The reversed dynamo phenomenon is a property of large $\Pm$.
Since $\Pm\propto T^4/\rho$ \citep[e.g.,][]{BS05},
where $T$ is the temperature, large $\Pm$ tend to
occur in stellar coronae and in galaxies where $\rho$ is very small.
Although dynamo action is always possible when $\Rm$
is large, it can only occur at LS, that is, for $k<k_\ast$,
because at SS, or for $k>k_\ast$, reversed dynamo action prevails.
This implies that most of the magnetic energy will be returned into
kinetic energy and then dissipated viscously.

It is not necessarily the case
that stellar coronae are dynamos since they are primarily driven through
magnetic stresses that build up in response to photospheric footpoint
motions, i.e., the energy is injected into the magnetic energy reservoir and
does not require a conversion from kinetic energy through a dynamo process.
This difference is highlighted in \Fig{Powerspec_Transfer_SSD_Corona} where we present
quantities similar to those in \Fig{pspec_lor_kin} for the LES dynamo setup with
$\Pm=54.6$ and a corona setup from \citet{Rem17} with $\Pm=39$. The corona is evaluated
in the height range of $12-24.8$~Mm above the photosphere. In the dynamo case,
$T_{\rm Str}$ is positive on almost all scales. $T_{\rm Adv}$ is negative on large and
positive on small scales, implying a transport of the magnetic energy induced by
$T_{\rm Str}$ to small scales, where it is dissipated most effectively. In the corona
case, $T_{\rm Adv}$ is positive on all scales since the corona is driven through the
Poynting flux generated by magnetoconvection in the photosphere. $T_{\rm Adv}$ is
balanced by a mostly negative $T_{\rm Str}$ (no dynamo). The energy is transferred to kinetic
energy through a $T_u$, which is positive on all scales, except for the smallest wavenumber.
While the dynamo case does have the requirement $T_u^- > T_u^+$, this is not the case
for the externally driven corona.
The scaling shown in \Figp{pspec_lor_posneg}{b} is then no longer valid.
Indeed, coronal transfer functions for smaller $\Pm$ look
qualitatively similar to the high $\Pm$ case presented here, i.e., we do not find
the $\Pm$ dependence of $k_*$ in this study. However, in a corona setup
with a 4 times larger domain, we found on large scales a more extended region with
$T_u < 0$. In spite of the significant differences in the underlying transfers,
\citet{Rem17} found that even for the corona, $\Pm$ determines the value of
the ratio $\epsK/\epsM$, with dominance of viscous dissipation in the high
$\Pm$ regime. Specifically, the simulations of \citet{Rem17} that use
the coronal arcade setup have $\epsK/\epsM$ ratios of $0.23$,
$0.83$, and $6.6$ for $\Pm=0.23$, $2.3$, and $39$, respectively (averaged over
the height range of $12-24.8$~Mm above the photosphere).
The transfer functions for the latter case are
presented in \Fig{Powerspec_Transfer_SSD_Corona}.

It appears that the dependence of $\epsK/\epsM$ on $\Pm$ is a more fundamental
property of MHD, whereas the $\Pm$ dependence of $k_*$ is specific to dynamo setups
that do have, in addition, the constraint $T_u^- > T_u^+$.
Furthermore, low density plasmas with large values of $\Pm$ are only
weakly collisional, and therefore departures from Maxwellian
particle velocity distributions and other kinetic effects are
expected to play a role \citep{Scheko09}.
Much of the kinetic energy may therefore go directly into particle
acceleration.

\section{Conclusions}

Our work has highlighted a qualitatively new feature of dynamos at large
magnetic Prandtl numbers, namely the conversion of magnetic energy back
into kinetic energy at high wavenumbers or SS.
This corresponds to reversed dynamo action.
It is responsible for the dissipation of energy through viscous heating
rather than through very narrow current sheets that are traditionally
thought to be responsible for energy dissipation in the corona
\citep{MBS88,GN96}.
Current sheets do also occur in large magnetic Prandtl number simulations,
but they are not directly responsible for dissipating significant
amounts of energy.
Instead, viscous energy dissipation is found to be the main mechanism
for liberating energy.
Viscous energy dissipation is a consequence of positive Lorentz force
work, which does occur in proximity of current sheets.
While the resulting plasma flows can transport kinetic energy away from
current sheets, the resulting viscous dissipation may still happen in
close proximity of current sheets, in particular when $\Pm$ is large,
and could therefore remain indirectly connected with them.

In kinetic simulations such as those of \cite{Rincon16} and \cite{ZWUB17},
reversed dynamo action may be chiefly responsible for particle acceleration.
Indeed, the same work term that leads to the conversion of magnetic
into kinetic energy also characterizes first-order acceleration of non-thermal
particles by curvature drift \citep{BL16}.
It will therefore be important to verify this aspect of particle
energization in future studies using kinetic simulations.
A direct comparison with our work is hampered by the fact that the concept
of a magnetic Prandtl number does not exist in collisionless plasmas.
A natural question would then be, whether $T^4/\rho$ can be regarded as
its proxy, as one would expect if Spitzer theory were applicable.

\acknowledgments
We thank Andrey Beresnyak for useful discussions in the beginning of this
work, and are grateful to Mausumi Dikpati and Kandaswamy Subramanian,
as well as an anonymous referee, for comments on the paper.
This research was supported in part by the Astronomy and Astrophysics
Grants Program of the National Science Foundation (grant 1615100),
and the University of Colorado through
its support of the George Ellery Hale visiting faculty appointment.
The National Center for Atmospheric Research is sponsored by the
National Science Foundation.
We acknowledge the allocation of computing resources provided by the
Swedish National Allocations Committee at the Center for Parallel
Computers at the Royal Institute of Technology in Stockholm
and the use of computational resources
(doi:10.5065/D6RX99HX) at the NCAR-Wyoming Supercomputing Center
provided by the National Science Foundation and the State of Wyoming,
and supported by NCAR's Computational and Information Systems
Laboratory.


\end{document}